\documentclass[twocolumn,preprintnumbers,amsmath,amssymb]{revtex4}
\usepackage{graphicx}
\usepackage{dcolumn}
\usepackage{bm}

\begin{document}
\title{Soliton fractional charges  in graphene nanoribbon and polyacetylene: similarities and differences   }
\author{S.-R. Eric Yang
\footnote{Corresponding author, eyang812@gmail.com}}
\affiliation{Anam-dong 5, Seongbuk-gu, Department of Physics, Korea  University, Seoul,
Korea, 02481\\}

\begin{abstract}
An introductory overview  of  current research developments  regarding  solitons and fractional  boundary charges in graphene nanoribbons is presented.   Graphene nanoribbons and polyacetylene have chiral symmetry and  share numerous similar properties, e.g., the bulk-edge correspondence between the Zak phase and the existence of edge states, along with  the presence of chiral boundary states, which  are important for charge fractionalization.
In polyacetylene, a fermion mass potential in the Dirac equation produces an excitation gap, and a twist in this scalar potential produces a zero-energy chiral soliton.  Similarly, in a gapful armchair graphene nanoribbon,   a distortion in the chiral gauge field can produce  soliton states.  
In polyacetylene,  a  soliton is bound to  a domain wall connecting two different dimerized phases. In    graphene nanoribbons, a  domain-wall soliton connects two topological zigzag edges with different chiralities. However, such a soliton does not display spin-charge separation. The existence of a soliton in  finite-length polyacetylene can induce formation of fractional charges on the opposite ends. In contrast, for gapful graphene nanoribbons, the antiferromagnetic coupling between the opposite zigzag edges induces integer boundary charges.   
The presence of disorder in graphene nanoribbons partly mitigates antiferromagnetic coupling effect.   Hence, the average edge charge of  gap states with energies within a small  interval is $e/2$, with significant charge fluctuations.   However, 
midgap states exhibit a well-defined charge fractionalization between the opposite zigzag edges in the weak-disorder regime.   Numerous   occupied soliton states  in a  disorder-free and doped zigzag graphene nanoribbon  form      a solitonic phase.

Keywords: chiral symmetry, fractional charge, topological insulator, soliton, graphene nanoribbon

\end{abstract}

\maketitle

\section{Introduction}

Graphene has considerable potential,  not only for  spintronic applications, but also for  fundamental physics \cite{Nov,Zhang}. In particular, graphene systems  have topologically protected chiral zigzag edge modes \cite{Wak,Ryu}.  An excellent opportunity   to observe these  boundary charges has recently arisen, as rapid progress has been made in the fabrication of atomically precise graphene nanoribbons (GNRs) \cite{Cai,Hao}.  Thus, the generation mechanism of solitons/vortices is at the center of research on  graphene materials \cite{Pac}.

\begin{figure}[!hbpt]
\begin{center}
\includegraphics[width=0.5\textwidth]{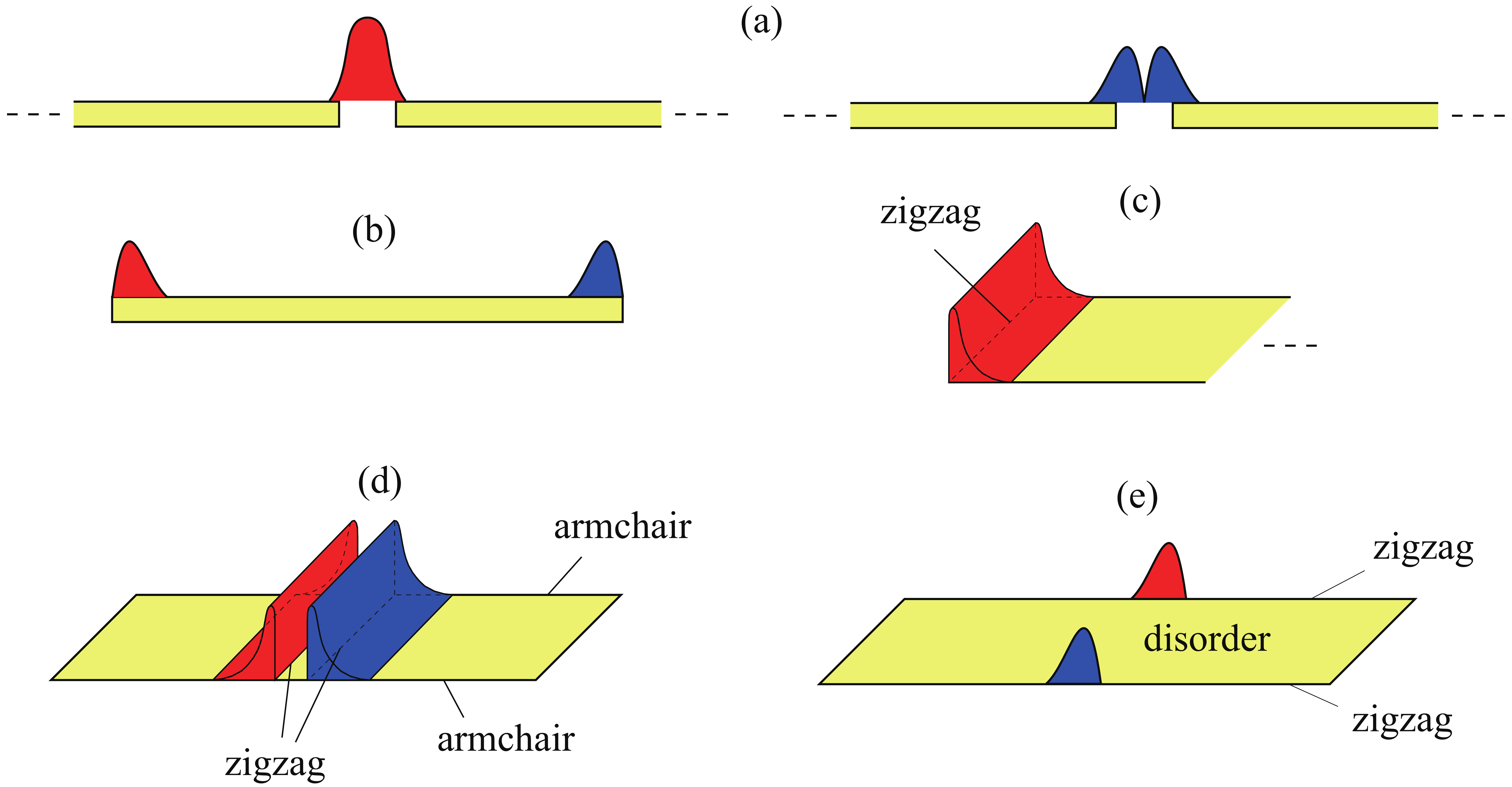}
\end{center}
\caption{Possible soliton states discussed in this paper.  The probability densities of a soliton on A and B carbon atoms are  indicated by red and blue colors, respectively.  (a) Chiral zero modes  of polyacetylene: the domain-wall soliton (left) and antisoliton (right) have different chiralties. (b) End-state soliton of polyacetylene with charge fractionalization: this soliton can be represented by  the bonding or antibonding linear combination of the chiral end states.  (c) Chiral zero mode on zigzag edge of semi-infinte armchair graphene ribbon.    (d) Domain-wall soliton of  armchair graphene nanoribbon  under local tensile strain. The probability densities on the A and B carbon atoms have a  small overlap.    (e) Charge fractionalization on zigzag edge of interacting  disordered zigzag nanoribbon. The probability densities   on the A and B carbon atoms are well separated.}
\label{soliton}
\end{figure}

 An object of one-dimensional insulators qualifies  as a soliton if   it  derives half its fractional spectral weight from each of the conduction and valence bands.   Jackiw and Rebbi \cite{Jack1} showed that a twist in the scalar mass potential in the Dirac equation,   connecting two degenerate groundstates, can produce a zero-energy soliton  state.
Furthermore, the independent work of Su, Schrieffer, and Heeger \cite{Su0,Sol} showed the presence of  a soliton (kink) in polyacetylene.  This kink exists between two different dimerized  phases and is called  a domain-wall soliton (see Fig.\ref{soliton}(a)).  A domain wall supports either a soliton or an antisoliton, but not both. A many-body ground state with the  solitonic state   unoccupied by an electron  has  an unusual charge  $Q$ and spin $S$ relation: $Q=e$ and $S=0$ (here, the contribution from the  positive ion background charge is included; $e>0$ is the elementary charge).  When the solitonic state is occupied, the following values are obtained:
$Q=0$ and $S=1/2$ .   
In finite-length polyacetylene, a soliton may exist with fractional {\it boundary} charges at the two end points (one for each end), as shown in Figure\ref{soliton} (b).

A domain-wall soliton can also exist in a GNR \cite{Sak,Jeong} (see Fig.\ref{soliton} (d)).  A single domain wall   in a GNR  can support  both a soliton and antisoliton, in contrast to polyacetylene; consequently, no unusual spin and charge relation holds.  In a doped zigzag GNR (ZGNR), numerous    soliton states  can be occupied, forming      a solitonic phase \cite{Luis}.
However,  no   fractional {\it boundary} charges are found in a GNR.   This is due to   {\it  antiferromagnetic} coupling between the zigzag edges \cite{Wak}.  
In the presence of disorder, undoped GNRs with zigzag edges   form a Mott-Anderson insulator,  and  the effect of antiferromagnetic coupling between the opposite zigzag edges is partly mitigated.  Hence,  in the weak--disorder regime, a  midgap state  can have $e/2$ fractional boundary charges on the opposite zigzag edges, i.e., one for each edge \cite{Jeong2}, as shown in Fig.\ref{soliton} (e) (zigzag edges  are located along the ribbon direction in ZGNRs and not at the end points).   Disorder acts as a singular perturbation on zigzag edge states. This is a consequence of a non-trivial interplay between disorder and electron interactions.    Note that an interacting disordered ZGNR has also non-trivial localization properties: the gap states  are localized  whereas    the states outside the gap are  delocalized, and the usual  localization theory does not apply.    

In this work,  the  similarities and differences  between solitons in GNRs  and polyacetylene are explained in detail. Various solitons of GNRs and polyacetylene are shown schematically in Fig.\ref{soliton}.   One important common feature is the chiral symmetry, which guarantees  that topological edge states exist in both systems.   This feature   is manifested  by the bulk-edge correspondence between the Zak phase and the existence of edge states.  The {\it chiral boundary modes} are shown to be intimately related to the fractional boundary charges, which  are formed from mixed chiral modes of   bonding or antibonding linear combinations of chiral modes.   The main differences between GNRs  and polyacetylene are "fermion doubling" and  the role of electron-electron interactions in charge fractionalization, as explained below.   It is hoped that this introductory overview of solitons will improve  understanding of  fractional boundary charges and  stimulate experimental searches for     $e/2$ charges  in GNRs.  Detection of these fractional charges may facilitate the study  of particles that obey fractional statistics in ZGNRs.

This paper is organized as follows.
In Sec. II, the connection between the boundary modes and the Zak phase (bulk-edge correspondence), which is based on chiral symmetry, is explained.
In Sec. III, a short review of the  formation of domain-wall  and end solitons with fractional end charges  in polyacetylene is provided.  
Domain-wall and end solitons of GNRs are reviewed
in Sec. IV.    The formation of fractional boundary charges in ZGNRs is more complex  than that in   polyacetylene, because of the  antiferromagnetic coupling between the well-separated  zigzag edges.  In Sec. V, we argue that   disorder  partially mitigates the effect of the edge antiferromagnetism and induces formation of fractional charges of the midgap states.   The conclusion is presented in Sec. VI.

\section{Chiral symmetry, Zak phase, and boundary modes  }

Before describing the topological properties of solitons in polyacetylene and GNRs, the importance of the chiral symmetry of these systems is explained.
The presence of    topological boundary 
states in GNRs and polyacetylene  is intimately related to  the presence of  chiral symmetry (sublattice symmetry) \cite{Ryu}. 
This behavior yields   {\it bulk edge correspondence}, for which   an edge  mode must exist when the bulk topological Zak phase is $\pi$.

\subsection{Chiral symmetry }

Polyacetylene and GNRs have two inequivalent sublattices (here labeled  A and B).    
Under  the chiral operation $\Gamma$, the site  annihilation operators  transform according to $a_{iA}\rightarrow a_{iA}$  and  $a_{iB}\rightarrow -a_{iB}$, where the index $i$ denotes a  unit cell containing  one A and one B carbon atom.  The  nearest-neighbor tight-binding Hamiltonians of
polyacetylene and GNRs satisfy the  anticommutation relation $ \{\Gamma,H \}=0$.    This symmetry is not exact when next nearest-neighbor hopping is included.  However, it yields a massless Dirac equation, which well describes low-energy excitations.
Single-particle eigenstates with non-zero eigenenergies $E$ and $-E$ are related by the chiral operation   $\Gamma|{ \psi_E}\rangle=|\psi_{-E}\rangle$.   Thus, chiral symmetry implies particle-hole symmetry.
The probability densities of $|{ \psi_E}\rangle$ and $\Gamma|{ \psi_E}\rangle$ are identical.  Of special interest are the single-particle eigenstates of the chiral operator, which  represent  chiral  topological states.

\subsection{Zak phase and edge charge}

One can show that the 
presence of a  boundary charge in a system with chiral symmetry can be related to
a bulk topological number (see the next subsection).     This number is
the Zak phase.  In this subsection  the Zak phase and its relation to the edge charge are discussed.
The Zak phase of a one-dimensional band structure is a topological Berry phase acquired by an electron as it moves adiabatically through the first Brillouin zone \cite{Zak} 
\begin{eqnarray}
Z_{1D}= \sum_{l\in occ}i\oint_{\text{B.Z.}}\left\langle u_{lk}|
\nabla_{k}u_{lk}\right\rangle dk,
\label{Zak}
\end{eqnarray}
where $\left|u_{lk}\right\rangle$ is the periodic part of the Bloch
wave function of the $l$th band  and the sum is over the occupied bands.   Note that $Z_{1D}$ is a {\it bulk} property  determined by the band structure of a periodic system.  Even when the chiral symmetry is broken, a one-dimensional periodic insulator with inversion/reflection  symmetry has   $Z_{1D}$ equal to  either $0$ or $\pi$ mod $2\pi$ \cite{Zak} (the modular $2\pi$ is a consequence of the gauge invariance).   

According to the modern theory of polarization, $Z_{1D}$ is related to the bulk polarization  of a periodic system \cite{Van,Van1}, such that 
\begin{eqnarray}
P=\frac{e}{2\pi}Z_{1D}. \nonumber\\
\label{Zak11}
\end{eqnarray}
Note that, for one-dimensional polarization, $P$ is defined as the dipole moment per length.  Consider   an {\it insulating} edge of the finite length system generated by cutting the periodic system.    The Zak phase  of the periodic system can be  related to the magnitude of the edge charge $Q$ \cite{Van}, such that  
\begin{eqnarray}
Q=\vec{P}\cdot \hat{n}\ .  
\label{Zak22}
\end{eqnarray}
This is an example of a bulk edge correspondence.  Note that $Q$ is located at the edge and the  direction of  $\hat{n}$ is  perpendicular to the edge.  
A   {\it rectangular} armchair GNR (AGNR) has two long armchair edges and two short zigzag edges.  The possible values of the  zigzag edge charge  
may be computed using  the
 $Z_{1D}$ of the periodic AGNR: $Z_{1D}=2\pi N$, where $N$ is an integer \cite{Jeong1}.   According to   (\ref{Zak11}) and   (\ref{Zak22}),  the edge charge is an {\it integer} with  $Q=eN$.        
Note that one cannot compute the possible charge values  on the zigzag edges of a ZGNR using Eq.(\ref{Zak22}), as these edges are parallel to  the ribbon direction.  As $Z_{1D}$ is multi-valued  Eq.(\ref{Zak22}) gives  possible values of $Q$ only.   The actual value of $Q$ must be computed  with consideration of the coupling between   the edge   and bulk.  The computed values for polyacetylene and GNRs are given in subsections III B and IV B, respectively.

\subsection{Zak phase of polyacetylene }

The chiral symmetry of polyacetylene  yields a non-trivial
topological  Zak phase \cite{Ryu}.   It is instructive to   explicitly compute the Zak phase of polyacetylene and to relate it to a fractional end charge.
Consider an infinitely long polyacetylene specimen. 
  The tight-binding Hamiltonian is 
\begin{eqnarray}
H=-t \vec{g}(k)\cdot \vec{\sigma},
\end{eqnarray}
and the chiral operator is $\Gamma=\sigma_z$. Chiral symmetry means   $ \{\sigma_z,H \}=0$.   This  anticommutation relation implies  that $\sigma_z$ is absent from $H$.   This point is crucial as it implies that  the vector $\vec{g}(k)$ does not have a z-component\cite{Kane}.  It can be written as 
\begin{eqnarray}
\vec{g}(k)=|\rho(k)|\left(  \begin{array}{l}
cos\phi(k) \\
sin\phi(k) \\
 \end{array} \right),
 \label{wavef}
\end{eqnarray}
where the phase $\phi(k)$ satisfies
$cot \phi(k) = \frac{t'/t}{sin ka_0} + cot ka_0$ and 
$\rho(k)= t'/t + e^{-ika_0} $. Here, $t'$ and $t$ are, respectively, the  intra and inter cell hopping parameters (see Fig.\ref{dimerb}).   The unit  cell length is $a_0$.   The eigenvectors are
\begin{eqnarray}
u(k)=\frac{1}{\sqrt{2}}\left(  \begin{array}{l}
e^{-i\phi(k)} \\
\pm 1 \\
 \end{array} \right)
 \label{eigen}
\end{eqnarray}
and the eigenvalues are $\pm t |\vec{g}(k)|$.
The pseudospin of an eigenstate is defined as the expectation value of $\vec{\sigma}$
\begin{eqnarray}
\vec{\Sigma}(k)=\left<\vec{\sigma}\right>=\left(  \begin{array}{l}
cos\phi(k) \\
 sin\phi(k) \\
 \end{array} \right).
 \label{pseudo}
\end{eqnarray}
As $k$  varies across  the
Brillouin zone, the vector $\vec{\Sigma}(k)$   rotates on a circle.  The Zak phase is related to the rotation angle  $\phi(k)$ of $\vec{\Sigma}(k)$, where 
\begin{eqnarray}
Z_{1D}=\frac{1}{2}\oint_{\text{B.Z.}}  \frac{\partial \phi(k)}{dk} dk.   
\label{phase}
\end{eqnarray}
As $k$  varies across the
Brillouin zone, $Z_{1D}$ is given by the difference $\phi(\pi/a)-\phi(-\pi/a)$.
When the unit cell of a periodic polyacetylene specimen contains   a long bond ($t'/t < 1$ ), $Z_{1D}$ is $\pi$ mod $2\pi$.  This reflects the fact that the bulk polarization can depend on the choice of  unit cell \cite{Van1}.     
In the case of a short-bond unit cell ($t'/t > 1$ ), $Z_{1D}$ is $0$ mod $2\pi$.  
A topological phase transition exists when the intra- and inter-cell tunneling coefficients are equal, i.e.,  $t'=t$ \cite{Kane}.  
In that case, long-bond polyacetylene is topologically non-trivial  whereas short-bond polyacetylene is topologically trivial. It can be demonstrated that the non-trivial value of $Z_{1D}$ induces the emergence of edge states in a finite system with open boundary conditions \cite{Del}.  Such a soliton has an end charge of $e/2$ (See Fig.\ref{soliton} (c)).  This fractional value can also be derived directly from Eqs. (\ref{Zak11}) and (\ref{Zak22}).

\subsection{Edge modes of graphene sheet and Zak phase}

The bulk-edge correspondence also holds for a semi-inifinte  graphene sheet, for which the edge may be a zigzag or armchair.  The number of  edge modes with wavevector $k_{\parallel}$ parallel to the edge direction can be determined using the  bulk-edge correspondence.
For a two-dimensional band structure, one can  define an analog of the one-dimensional  Zak phase.    For  a wavevector $k_{\parallel}$,  this  Zak phase is defined as
\begin{eqnarray}
Z_{2D}(k_{\parallel})=i\oint_{\text{B.Z.}} \left\langle u_{\vec{k}}|
\nabla_{k_{\perp}}u_{\vec{k}}\right\rangle dk_{\perp},   
\label{Zak2}
\end{eqnarray}
where the wavevector    $k_{\perp}$ is perpendicular to $k_{\parallel}$.   Integration over $k_{\perp}$ should be performed on a cut of a 2D Brillouin zone in a direction transverse to $k_{\parallel}$. One can show that $Z_{2D}(k_{\parallel})/\pi$ gives the number of  edge states  at 
$k_{\parallel}$ (this is an example of the bulk-edge correspondence \cite{Ryu,Del}).
Then, 
\begin{eqnarray}
Z_{2D}(k_{\parallel})=\frac{1}{2}\oint_{\text{B.Z.}}  \frac{\partial \phi(\vec{k})}{dk_{\perp}} dk_{\perp}.   
\label{phase}
\end{eqnarray}

When the integration is performed    along  the direction of an armchair edge,  $Z_{2D}(k_{\parallel})$  is of $\pi$; however, it is zero when  integrated in the direction of a  zigzag ribbon.  This implies the existence of  {\it chiral} edge states  on a zigzag boundary, but not on an armchair edge \cite{Ryu,Del}.  Tight-binding calculations confirm this result (see subsection IV B).  Thus, these edge states are protected topologically.  Note that the boundary charge given by Eq.(\ref{Zak22}) cannot be calculated from  $Z_{2D}(k_{\parallel})$; one must use $Z_{1D}$ instead.  
There is no reason why  $Z_{1D}$ should be identical to  $Z_{2D}(k_{\parallel})$.

\section{Solitons  in polyacetylene}

Topological edge modes are soliton modes.    Some salient features of solitons   in polyacetylene are briefly reviewed here (see Ref.  \cite{Sol} for a comprehensive review; refs. \cite{Kane, GirvinKun} also give a nice overview of solitons).

\subsection{Domain-wall soliton  in polyacetylene}

\begin{figure}[!hbpt]
\begin{center}
\includegraphics[width=0.4\textwidth]{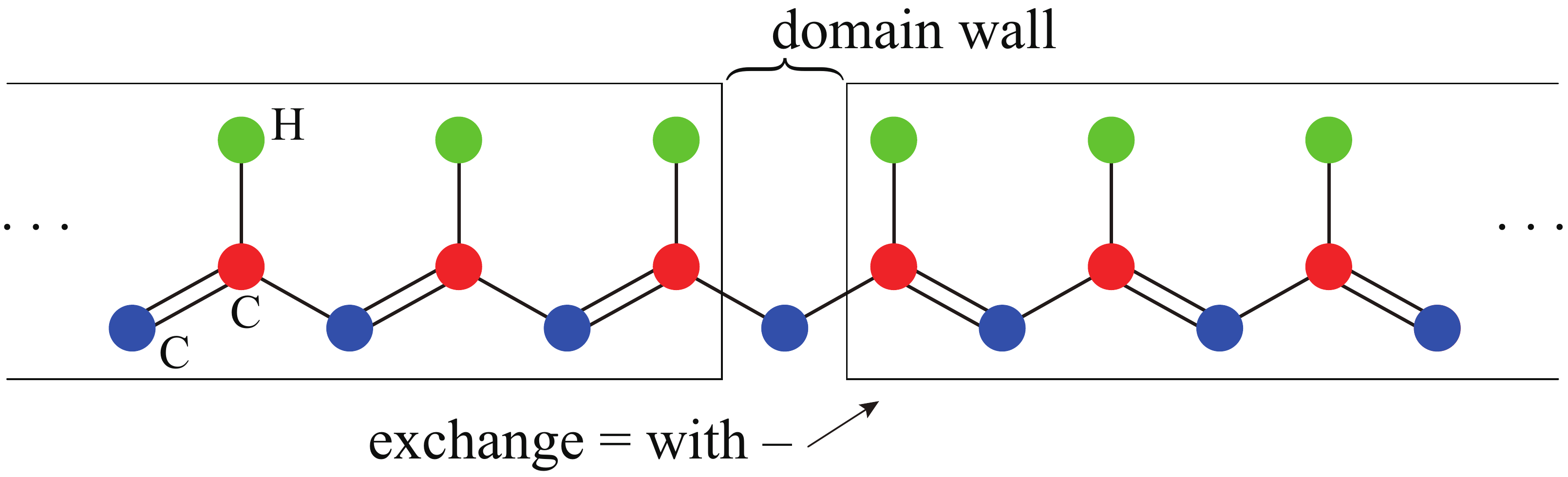}
\end{center}
\caption{Two semi-infinite  polyacetylene chains with a domain wall, which connects two different dimerized phases labeled  $-m$ and $m$.  
A double bond is shorter  than a long bond and a Peierls lattice distortion is present.  The directions of the double bonds vary in these two phases.   Periodic boundary conditions cannot be imposed on this system as there is only one domain wall.   Either a soliton or antisoliton is possible, but not both.   }
\label{dimera}
\end{figure}

\begin{figure}[!hbpt]
\begin{center}
\includegraphics[width=0.5\textwidth]{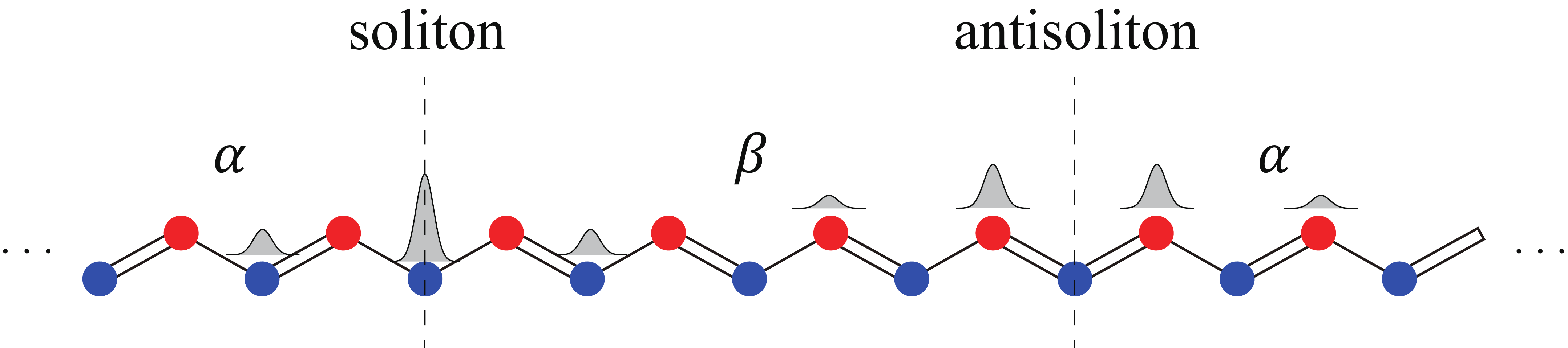}
\end{center}
\caption{In a periodic system with two domain walls, a soliton-antisoliton pair must exist.   Their site probability densities are  shown.   For a soliton (antisoliton), the probability density is finite on A (B) carbon atoms only.}
\label{periodic}
\end{figure}

Consider two semi-infinite    polyacetylene specimens. 
A domain wall  connects two degenerate ground states of  the polyacetylene, as shown in  Fig.\ref{dimera}.  
In the continuum description of this system, a fermion mass potential in the Dirac equation produces an excitation  gap, 
and a twist in this scalar potential produces a {\it zero}-energy soliton  and fermion fractionalization \cite{Jack1}.   The Dirac equation\cite{Kane}  has the form
\begin{eqnarray}
H=-iv\hbar\partial_x\sigma_x+m(x)v^2\sigma_y,
\label{Ham}
\end{eqnarray}
where $\{\sigma_x, \sigma_y\}$ are Pauli spin matrices and  $v$ is the characteristic velocity.  The second term is the twisted scalar mass  potential.  The twist  is $-m$ for $x<0$ and $m$ for $x>m$, where $m$ is a constant  (this  system is not periodic as the $\pm m$ terms represent two different dimerized phases).  This equation has {\it one} zero-energy soliton (kink)  mode $\psi_0(x)$, which  is    bound to the domain wall and decays exponentially.   Its first and second components give the probability amplitudes of finding the electron on A and B carbons, respecitvely.  
However, only the A-component of the wavefunction is non-zero (this solution corresponds to a soliton that is very well localized near the   A-carbon atom  at $x=0$, as shown in Fig.\ref{periodic}).   A  soliton  {\it is a chiral mode}  and is topologically  robust, as it originates from a twist in the variation of the dimerization $m(x)$.

The following discussion demonstrates that the conduction and valence bands each  contribute a fractional spectral weight of $1/2$ to  a soliton.
For simplicity,  {\it spinless} electrons in  disorder-free and  half filled polyacetylene are considered first.    The conduction and valence bands obey chiral  symmetry  so that, for opposite energies $E$ and $-E$, there are   identical wave functions: $\psi_E(x)=\psi_{-E}(x)$ (particle-hole symmetry follows from chiral symmetry; the Fermi energy is $E_F=0$).  
A spectral analysis is performed using the  local density of states of   the {\it translationally invariant} and  non-invariant systems,  $\rho^{0}(E,x)$ and
$\rho^{kink}(E,x)$, respectively.
As the total weight is conserved before and after the translational symmetry is broken, we have
\begin{eqnarray}
\sum_{E}\rho^{kink}(E,x)=\sum_{E}\rho^{0}(E,x).
\end{eqnarray}
 Before  a soliton is introduced  there is no zero-energy state; thus, $\sum_{E}\rho^{0}(E,x)=\sum_{E\neq 0}\rho^{0}(E,x)$.  After a kink is introduced, the total  density of states (DOS) at $x$ is $\sum_{E}\rho^{kink}(E,x)=|\psi_0(x)|^2+\sum_{E\neq 0}\rho^{kink}(E,x)$.   From this  and the closure property of the eigenstates $\psi_E(x)$
\begin{eqnarray}
\sum_{E}\rho(E,x)=\sum_E\psi^+_E(x) \psi_E(x)=1,
\label{one}
\end{eqnarray}
we find
\begin{eqnarray}
\sum_{E\neq 0}\rho^{0}(E,x)=|\psi_0(x)|^2+\sum_{E\neq 0}\rho^{kink}(E,x).
\end{eqnarray}
The   induced DOS excluding the  $E=0$ state is
\begin{eqnarray}
\sum_{E\neq 0}\delta\rho(E,x)=\sum_{E\neq 0}( \rho^{kink}(E,x)-\rho^{0}(E,x) ) =-|\psi_0(x)|^2.\nonumber\\
\end{eqnarray}
As the  conduction and valence bands are symmetric, the contribution from the {\it occupied} valence band states is
\begin{eqnarray}
\sum_{E< 0}\delta\rho(E,x)=-\frac{1}{2}|\psi_0(x)|^2.
\end{eqnarray}
Thus, the occupied valence band  contributes  a {\it fraction} of $1/2$ to the total spectral contribution of the solitonic state, while   the unoccupied  conduction band contributes  another half (one half of a state is  missing from the valence band  and the corresponding charge is assumed to be in the  vicinity of the soliton \cite{Sol}).

When the electron spin is considered, each spin-up and -down soliton gap state
takes half the spectral weight from the valence band.  Thus, if the spin-up and -down soliton states are both empty,  the  localized soliton has   $Q=e/2+e/2=e$ and   $S=0$ \cite{Su0,Sol} (here, a   charge is defined as a depletion or surplus in the many-body {\it ground state  density} including the positive background charge).     When  the spin-up  soliton state is occupied while the spin-down state is empty, $Q=-e+(e/2+e/2)=0$ and  $S=1/2$.  Therefore, these solitonic states  have  unusual charge and spin relations \cite{Heeger}.

Next, consider a {\it periodic} polyacetylene specimen having two domain walls, see Figs.\ref{periodic}.  It is instructive to consider tight-binding solutions.  They are a soliton and an antisoliton  solutions, as shown in  Fig.\ref{periodic}.  For infinitely long  polyacetylene,  the energy difference between a soliton and antisoliton vanishes.
When both a soliton and antisoliton are present, they must be located at  {\it different} positions along a periodic  ring: a   soliton connects two dimerized phases, $-m\rightarrow m$, whereas an antisoliton connects $m\rightarrow -m$, as shown in  Fig. \ref{periodic}.  A soliton and an antisoliton are  chiral modes with different chirality: a soliton has only the A-component wavefunction and  an antisoliton has only
the B-component wavefunction, as apparent from  Figs.\ref{soliton}(a) and \ref{periodic}.   An antisoliton   is also an  eigenstate of the  chiral operator.  The energy spectrum varies strongly in the vicinity of the kink/antikink.   The total number of electrons or states in the filled valence band in the vicinity of the kink/antikink decreases by precisely $1/2$ per spin (see the derivation  below).

\subsection{End solitons of polyacetylene}

\begin{figure}[!hbpt]
\begin{center}
\includegraphics[width=0.27\textwidth]{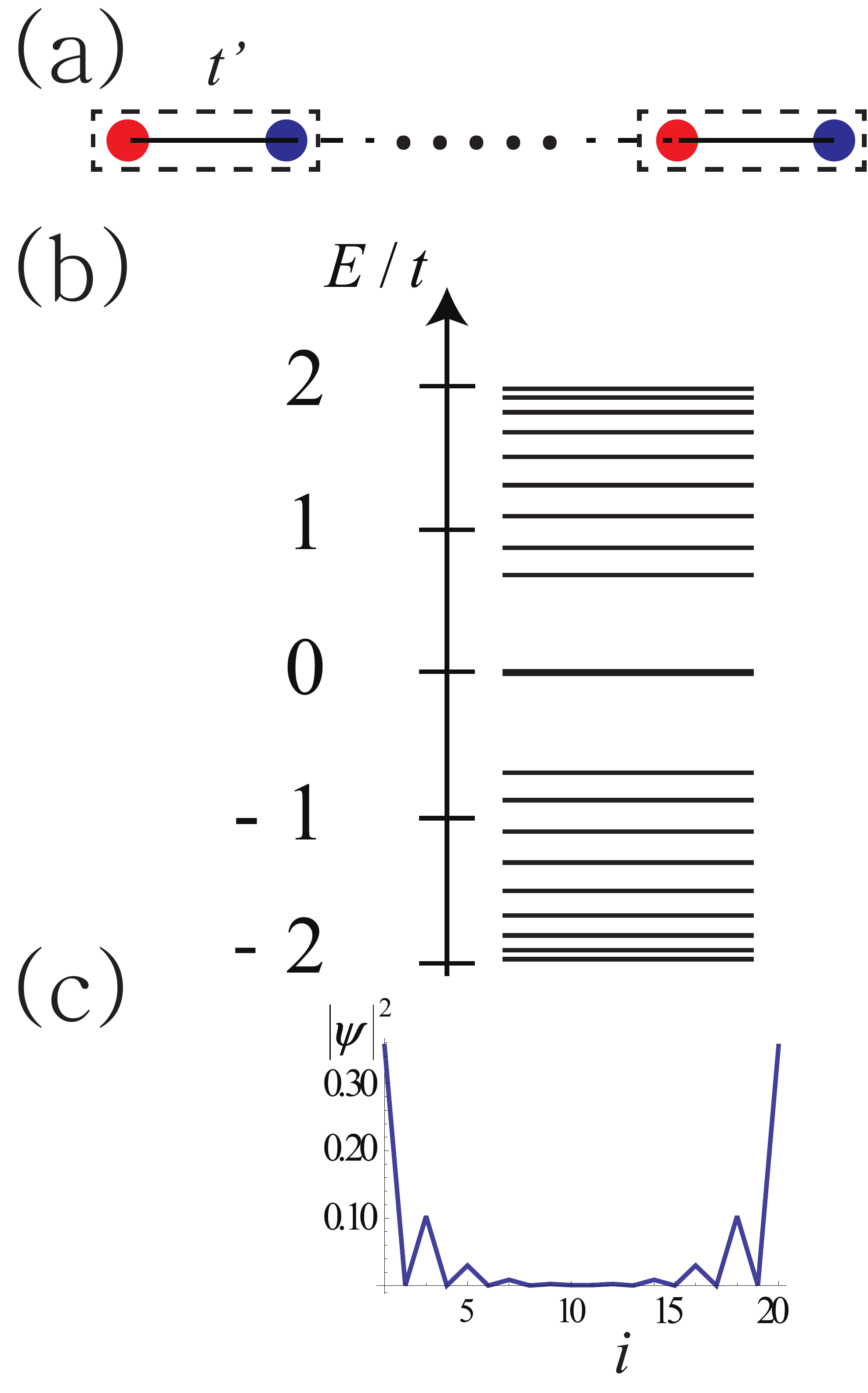}
\end{center}
\caption{(a)  Finite-length dimer chain with unit cell containing two carbon atoms connected by  single bond.   The Intra cell hopping  $t'$ is smaller than the inter cell hopping $t$.  (b) Tight-binding energy spectrum.   Two nearly  degenerate gap states  exist.  (c) Probability density of   gap state as  function of site index $i$.   A peak is apparent at the red (blue) site at the left (right) end. The probability densities of the bonding and antibonding states are almost identical.    }
\label{dimerb}
\end{figure}

\begin{figure}[!hbpt]
\begin{center}
\includegraphics[width=0.3\textwidth]{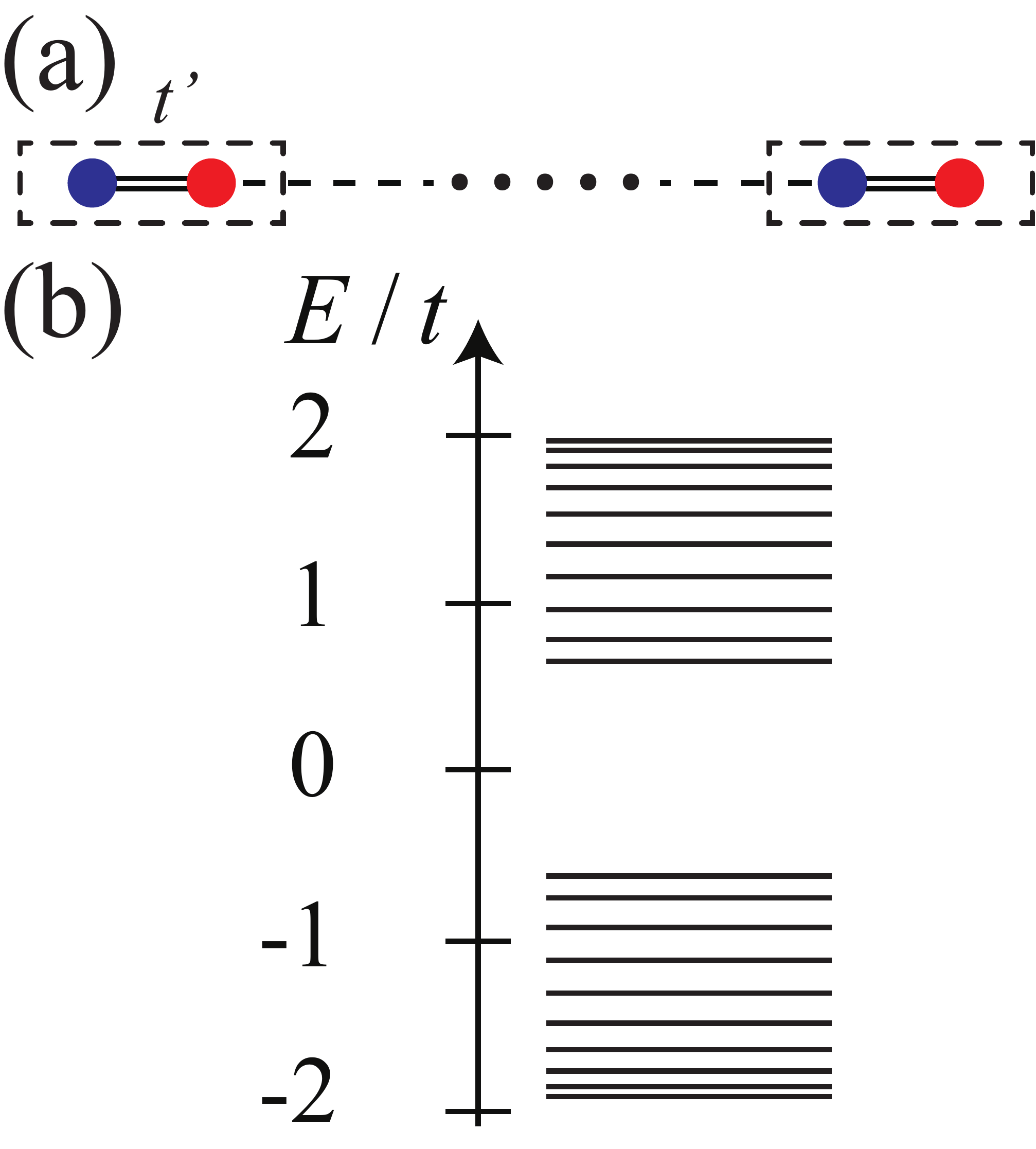}
\end{center}
\caption{ (a) Finite-length dimer chain with unit cell containing two carbon atoms connected by  double  bond.   The intra cell hopping  $t'$  is larger than the  inter cell hopping $t$.   Note that the unit cell is different from that shown in Fig.\ref{dimerb}. (b) Energy spectrum.  No gap states exist.}
\label{dimerb2}
\end{figure}

A soliton can also exist as a boundary charge.  Consider  {\it finite-length}  polyacetylene in one of the dimerized phases  (no domain wall exists as only one type of dimerized phase is present).  The electron density is uniform with occupation number   $n_i=1$ at all  sites $i$.    There are two types of {\it finite}  length polyacetylene, which have long or short end bonds, as shown in Figs.\ref{dimerb}  and \ref{dimerb2}, respectively.   

Tight-binding calculations show that, for the long-bond  unit cell shown in Fig.\ref{dimerb}(a),   two nearly degenerate bonding $\phi_B$ and antibonding  $\phi_A$ {\it gap} states exist with almost zero-energy (the energy splitting vanishes when the  system length becomes  infinitely large).  One half of the spectral weight of each of these  solitonic states is derived  from the conduction band, while the other half is from the valence band.      
As is $E_F=0$ , one state is occupied while the other is empty (see Fig.\ref{dimerb}(b)).      
The probability density of such a state   {\it splits} into two parts, located near the left and right end points (see Fig.\ref{dimerb}(c)).
If an electron is added   to  a solitonic state, the resulting 
electron density $\rho(x)$ has $-e/2 $ fractional charges near the {\it two ends} of the polyacetylene.   When an electron is removed, $e/2$ fractional charges appear near the ends.
Note that these solitons have  mixed chirality with different  chirality at the opposite ends (here referred to as {\it here refered to as mixed chiral states}). 
However,  their linear combinations $\phi_B\pm\phi_A$ are {\it chiral} and are located near either the left or right end points.   In the case of the short-bond unit cell, no end  state exists, i.e.,  no gap state exists,  as shown in  Fig.\ref{dimerb2}.    In periodic polyacetylene,  a topological phase transition occurs at $t=t'$ with a discontinuous change in the  value of the Zak phase  (see the discussion of the Zak phase and end charge in subsection II C).

\section{Solitons  in insulating GNR}

According to tight-binding calculations \cite{com}, an AGNR has a gap and is semiconducting when the transverse width is $L_y = (3M + 1)a_0$ or $3Ma_0$, where $a_0=\sqrt{3}a$ is the unit cell length of the graphene lattice, $a=1.42$ \AA\  is the carbon-carbon distance, and $M$ is an integer.  However, when $L_y = (3M + 2)a_0$, the AGNR  has no gap and is metallic (we do not consider this case here).     A rectangular  GNR  has two zigzag edges and two armchair edges.   When  the zigzag edges are longer than  the armchair edges, a ZGNR is realized.  In the opposite case, an AGNR is realized.  
 A periodic ZGNR with a bandstructure has only two zigzag edges and no armchair edges.

Graphene  lattice  is  two-dimensional, unlike that of polyacetylene. Chiral symmetry then leads to a phenomenon called "fermion doubling" \cite{FD} (K and K' valleys exist).  It implies that  a  GNR with the shortest width has {\it two} domain-wall states, in contrast to polyacetylene (for other values of the width the number of zero modes is an even integer \cite{Jeong}).   It should be also noted that a  soliton mode of a GNR that we describe below connects sites with different chiralities.   It can connect two  well-separated topological zigzag edges with opposite chiralities, and hence  it is topologically protected.   On the other hand, Sasaki et al. \cite{Sak} explored a graphene nanoribbon with a domain-wall soliton   connecting two distinct bonding structures (each structure has  a bond alternation similar to dimerization in  polyacetylene).


\subsection{Domain-wall soliton }

We describe  below a domain-wall soliton of a  semiconducting  AGNR under a local tensile strain\cite{Jeong}.  Such a domain-wall does not require    Kekul\'{e}-like  bond alternation.
Some of the salient features of the domain-wall soliton  are discussed in this subsection, and the similarities and differences in comparison to those of polyacetylene are delineated.

\begin{center}
\begin{figure}[!hbpt]
\includegraphics[width=0.35\textwidth]{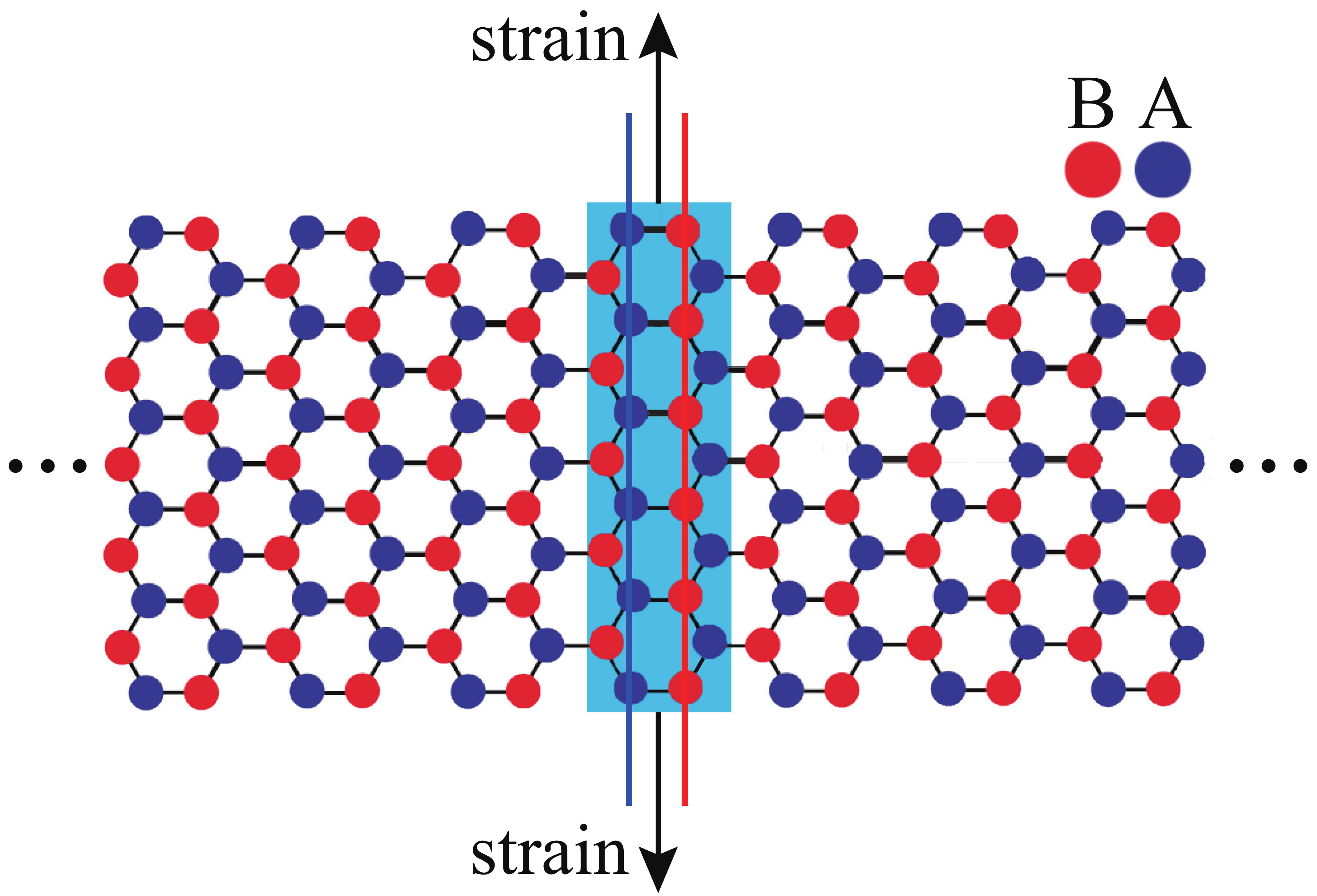}
\caption{AGNR with local tensile strain applied perpendicular to ribbon direction in region D  (blue area). A and B zigzag edge sites are shown on the blue and red lines.}
\label{tensile}
\end{figure}
\end{center}

Consider an infinitely long    semiconducting  AGNR.   Suppose we apply a local tensile strain perpendicular to the ribbon direction \cite{Jeong}, as shown in  Fig. \ref{tensile}. This induces changes in the hopping parameters between the carbon atoms in a rectangular  area $D$,  where strain is applied.  In   the continuum description of graphene,  such a distortion can be simulated by a chiral gauge field $\vec{A}_f(\vec{r})$ \cite{Sak}. For the $K$  valley, we have
\begin{eqnarray}
H_K=v_F\vec{\sigma}\cdot(   \vec{p}-\frac{e}{c}\vec{A}_f(\vec{r})   ),
\end{eqnarray}
where the Pauli spin matrices are $\vec{\sigma}=(\sigma_x,\sigma_y)$, $ \vec{p}$ is the momentum operator, and $v_F$ is the Fermi velocity of bulk graphene.  The second term is the chiral gauge vector field.  Similarly, for the $K'$ valley, we have
\begin{eqnarray} 
H_{K'}=v_F\vec{\sigma}'\cdot(\vec{p}+\frac{e}{c}\vec{A}_f(\vec{r})),
\end{eqnarray}
where $\vec{\sigma}'=(-\sigma_x,\sigma_y)$.  Note that the signs of the chiral gauge vector field differ in these equations.   The chiral vector is a constant in $D$,  its direction is along the x-axis, and it is zero outside   $D$.    However, it is not a real vector potential; rather, it  effectively describes the change in the hopping parameters in $D$, i.e., in  the domain wall.  The armchair edges {\it couple} the two valleys and the solutions are four-component wave functions \cite{com}.

\begin{center}
\begin{figure}[!hbpt]
\includegraphics[width=0.4\textwidth]{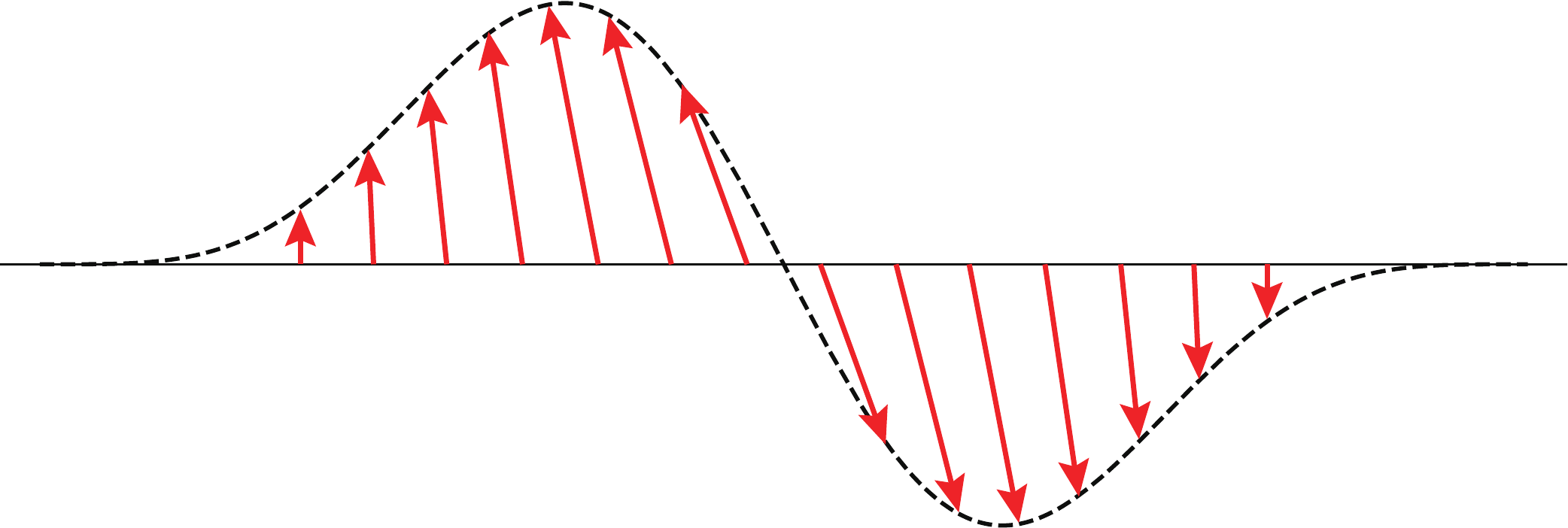}
\caption{Domain-wall soliton corresponds to a topological kink. Pseudospin vector rotates by an angle of $\pi$ out of the plane.}
\label{kink}
\end{figure}
\end{center}

For a ribbon with the shortest  width  these equations have {\it two} solutions with  $E\approx 0$, representing gap states, i.e., a domain wall supports {\it both}  solitonic and antisolitonic 
solutions with opposite energies \cite{Jeong}.  The antisoliton state is unoccupied at half filling.    {\it Half}  the spectral
weight of a soliton/antisoliton arises  from the conduction band and {\it the other half} from the
valence band.   The solitonic states  are {\it not} eigenstates of the chiral operator:    on the left and right hand sides of the domain, the  dominant chirality wavefunction is of  A- and B-type, respectively. In the left (right) part, the wave function is non-zero on A-type (B-type) zigzag edge sites, see Fig.\ref{tensile}. These states are mixed chiral states (see Fig.\ref{soliton} (d)).  However, a soliton and antisoliton  are connected to each other by chiral operation (they have opposite energy).   
When the tensile strain is sufficiently strong that the bonds in region $D$ almost break,   the left and right parts of the wavefunction have a rather small overlap.   
Note that the  pseudospin \cite{pseudo} of the solitonic state   rotates by $\pi$ across the domain wall,  indicative of a topological kink \cite{kink}, as shown in  Fig.\ref{kink}.    The crucial feature of a topological kink is that the total change is $\pi$ irrespective of  the manner in which the phase changes as the coordinate changes.
Unusual spin and charge relations are not expected in GNRs\cite{Jeong1}, as a {\it single}  domain wall can  simultaneously support both  a soliton and an antisoliton,   in contrast to  polyacetylene.  Note that, when both the soliton and antisoliton states are unoccupied, the valence band misses the total charge of $e$ per spin.

\begin{center}
\begin{figure}[!hbpt]
\includegraphics[width=0.4\textwidth]{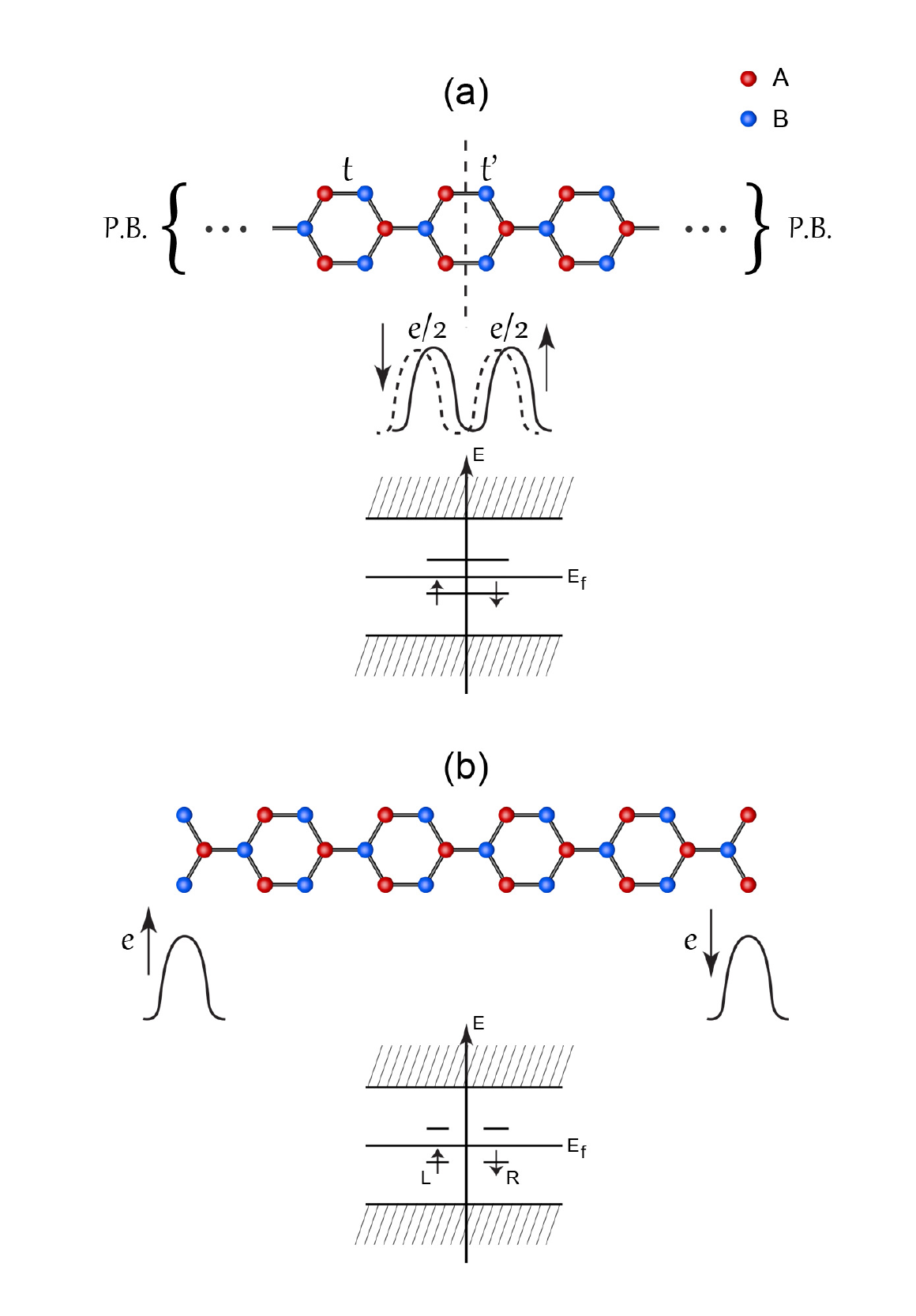}
\caption{(a) Periodic armchair graphene nanoribbon (AGNR) with  shortest width and modified
hopping $t'$ between two neighboring zigzag edges at center. P.B. represent periodic
boundary conditions.  A schematic illustration  of the probability densities of the  soliton gap states is also shown.  Some overlap between its   left and right probability densities occurs if
the hopping between these two edges, i.e., $t'$,  is non-zero.   
The energy spectrum of the gap states for a finite on-site electron repulsion $U$ is also shown.   In the limit of small $U$,  the energy splitting of the solitonic gap states vanishes and the  energies approach zero.
(b) In the limit $t'\rightarrow 0$  (corresponding to cutting of the bond), a rectangular AGNR is realized.  The solutions 
change qualitatively:  antiferromagnetically coupled integer charges  develop that are  localized on the left (L) and right (R) zigzag edges. }
\label{arm}
\end{figure}
\end{center}

The domain-wall soliton may be constructed  from a linear combination of chiral zigzag edges modes. 
It is instructive to analyze  this problem  using a simple  model \cite{Jeong}.     Suppose we consider a periodic  AGNR with a {\it short} width, and assume, for simplicity,  that only one bond is affected by tensile strain, as shown in  Fig.\ref{arm}(a).   When the distorted  hopping parameter vanishes, $t'= 0$, the bulk-edge correspondence indicates   that {\it chiral} modes $\phi_L$ and  $\phi_R$ develop on the left  and right  zigzag edges (see Fig.\ref{soliton}(c)).
When $t'\neq 0$, a domain soliton mode forms, which can be expressed as  a linear combination $\phi_L\pm\phi_R$ of  the chiral modes of the  left and right zigzag edges at $t'=0$, as shown in  Fig.\ref{arm}(a) (both bonding and antibonding combinations are possible).   An electron in  a  solitonic state   resides  near  the two  neighboring zigzag edges with {\it opposite chirality} \cite{Jeong}.    Its tight-binding probability density is divided equally between the left and right zigzag edges  (the solitonic wavefunction exhibits some overlap between the edges).  This domain  soliton mode is robust as  the edge chiral modes 
 $\phi_L$ and  $\phi_R$
are topologically protected \cite{Ryu,Del} (they  persist as long as the zigzag edges are not destroyed).
In the limit where   $t'\rightarrow 0$, a rectangular AGNR is realized, as shown in  Fig.\ref{arm}(b).   In addition, the effect of on-site repulsion is more significant  in comparison to its effect for $t'\neq 0$.  It makes the bonding and antibonding states $\phi_L\pm\phi_R$  no longer stable on the zigzag edges.  This is because on-site repulsion   induces antiferromagnetic coupling between  edge charges and mitigates the formation of fractional edge charges.

\subsection{Solitons of  interacting ZGNR}

\begin{center}
\begin{figure}[!hbpt]
\includegraphics[width=0.4\textwidth]{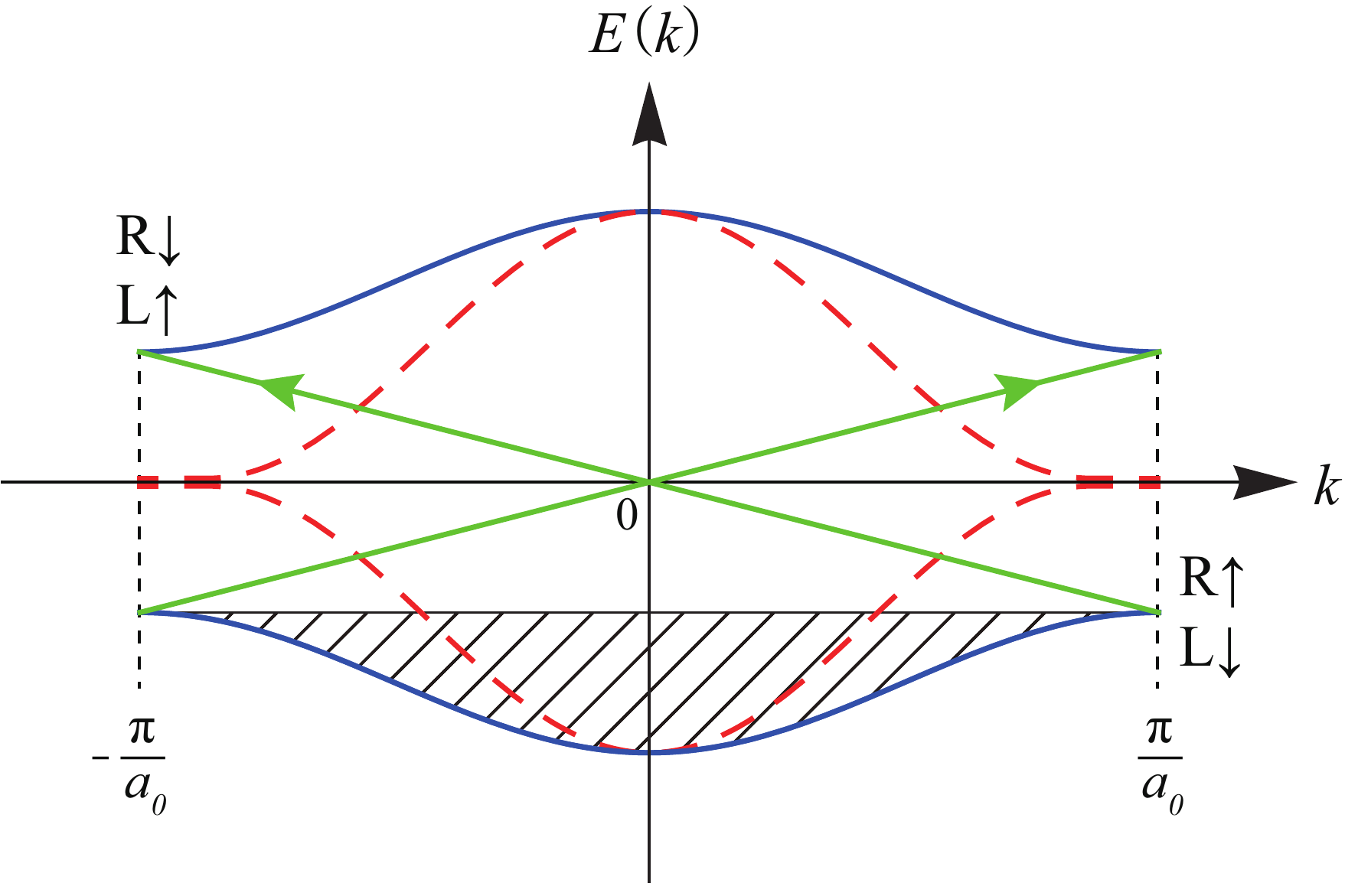}
\caption{Schematic display of band structure for  interacting ZGNR in absence of disorder (solid lines). (The gap size is exaggerated).  Only bands near the gap are displayed. The natures of the unoccupied and occupied    states near $k=\pm \frac{\pi}{a_0}$  are given  ($a_0$   is the unit cell length of
the ZGNR): $R$ and $L$ represent states localized on the right and left zigzag edges, respectively. Small arrows indicate spins.  The bandstructure for non-interacting electrons is represented by dashed lines.  There are numerous zero-energy states near the Brillouin zone boundary, which  split in the presence of on-site electron repulsion $U$ (this splitting is analogous to  the energy splitting shown in Fig.\ref{arm}.   The long arrow indicates coupling between states $R \uparrow$ and $L\uparrow$   or  $R \downarrow$ and $L\downarrow$.}
\label{Ubandst}
\end{figure}
\end{center}

In this subsection, end solitons of half-filled  periodic ZGNRs are considered.    The Hartree-Fock approximation \cite{Luis,Ross,Stau,Pis} (HFA) result  for such  a disorder-free system  displays  numerous pairs of occupied spin-up 
and -down chiral edge
states that are   located on the opposite zigzag edges.  They are solitons   of the type shown in Fig.\ref{soliton}(c) \cite{Wak}.   These states correspond to states near the first Brillouin zone boundary, as shown in Fig.\ref{Ubandst}.   Their number is even and increases linearly with the zigzag edge length  \cite{Jeong}.  Electron interaction is responsible for the excitation gap and edge antiferromagnetism \cite{Wak,Yang,Sor}.    The study of the Zak phase also suggests that the edge charge on a zigzag edge is an integer \cite{Jeong1}. 
There are no   well-separated solitonic boundary charges of $e/2$ on the zigzag edges, as    the antiferromagnetic coupling between the zigzag  edges produces integer charges.   

When a disorder-free ZGNR is doped, the additional electrons form a solitonic phase \cite{Luis} (then, $E_F>0$).  On the left zigzag edge, the edge spin profile exhibits a rotation of $\pi$ as the coordinate position varies from one end of the zigzag edge to the opposite end of the same zigzag edge, i.e., it rotates from spin-up to spin-down.  On the right zigzag edge, the edge spin profile rotates  from spin-down to spin-up.   The DOS develops  a sharp solitonic peak at $E=0$ in the middle of the gap.

\section{Soliton fractional charge of interacting disordered ZGNR}

 The domain-wall soliton of Fig.\ref{soliton} (d) has two fractional charges that overlap.  This is not a true charge fractionalization as the overlap between fractional charges  is not small.  As demonstrated in the previous section,  neither of the two zigzag edges of a rectangular AGNR  supports a fractional charge because of the edge antiferromagnetism, as shown in  Fig.\ref{arm}(b).
 
 An  impurity potential  or magnetic field can have a significant probability density redistribution effect  over   the opposite zigzag edges \cite{Park}.  For example,  when the reflection symmetry of a rectangular  AGNR is broken by a staggered potential,  the Zak phase is no longer quantized and the zigzag edge charge can take non-integer values \cite{Jeong1}.  In this case the Zak phase can  be related only approximately to the edge charge via Eq.(\ref{Zak22}).   Moreover, for some values of the strength of the staggered potential, numerous states of a rectangular AGNR are spin-split while the states  of the corresponding periodic AGNR are not.    For these GNRs    the Zak phase cannot be related to the edge charge.  

\begin{center}
\begin{figure}[!hbpt]
\includegraphics[width=0.5\textwidth]{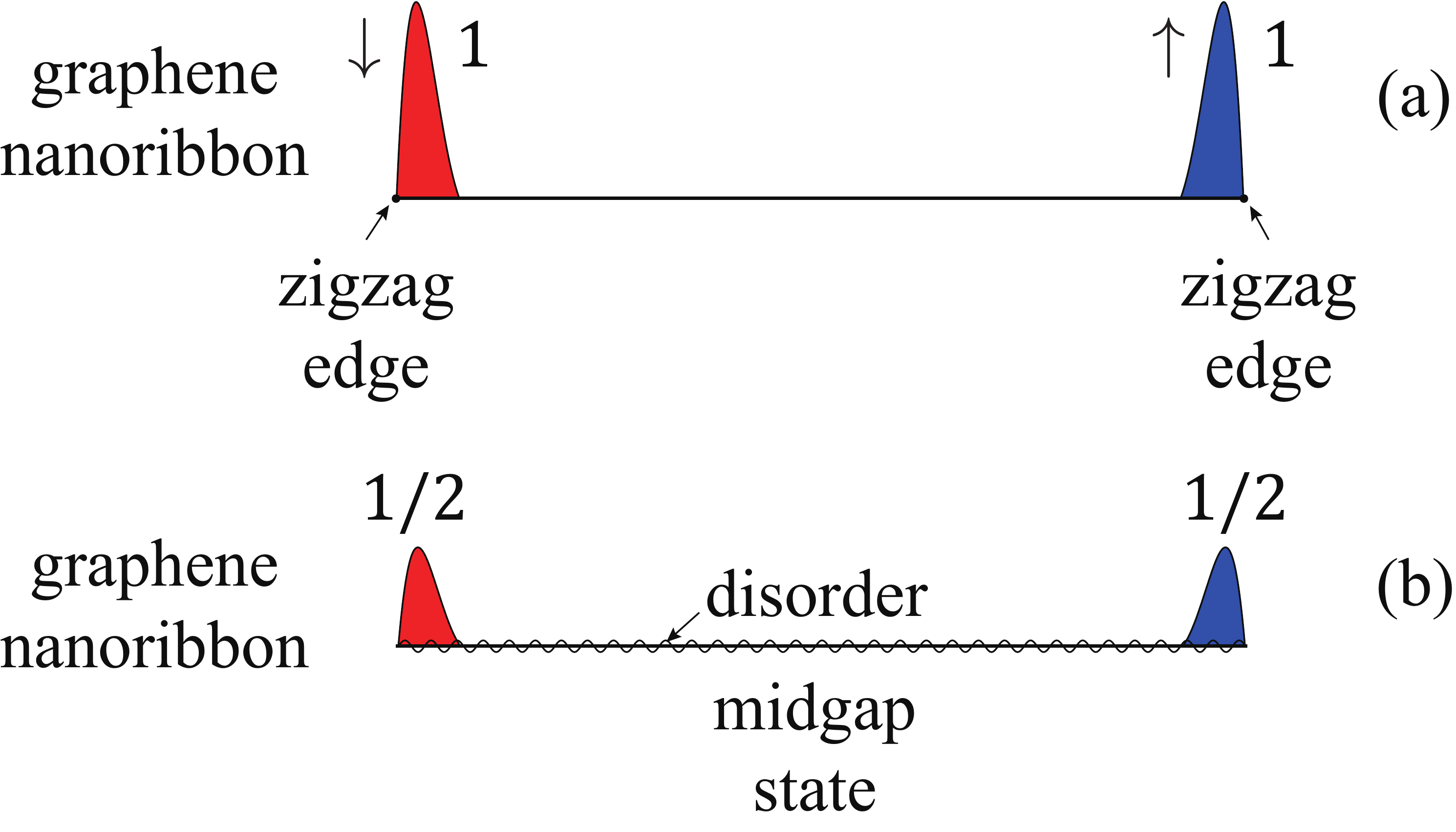}
\caption{(a)   In a clean and gapful ZGNR, the zigzag edges couple antiferromagnetically and no fractional boundary charge exists on them.   An edge state of this system displays an  integer boundary charge of one.  The shaded areas in the figure indicate the  probability densities of the edge states.  Note that the chirality differs among the zigzag edges.
(b)  In a disordered ZGNR the  antiferromagnetic coupling is partly weakened  and a  midgap state ($E\approx 0$) with a given spin value can display  fractional end charges of $e/2$.   Note that this state is  localized along the zigzag edges, in contrast to the  AGNR case shown in Fig.\ref{arm}(b).  The wavy line indicates a disorder potential.}
\label{av1}
\end{figure}
\end{center}

Let us examine the effect of disorder on charge fractionalization in  undoped interacting disordered ZGNRs.
{\it The gap of  a disordered ZGNR is  filled with localized  states, whereas     the states outside the gap are  delocalized} \cite{Jeong2}.   (The usual one-dimensional localization theory does not apply to GNRs \cite{Ando,Neto}).  A short-ranged disorder potential  induces stronger localization along the zigzag edges  than a long-ranged disorder potential \cite{Lima}.   
In addition,   spin-split states are also present \cite{Sor,Jeong1}, as in  a Mott-Anderson insulator \cite{Dob}.
 Let us analyze the scattering of the left and right edge  states  by a short-ranged disorder potential.   Consider
a spin-up electron at  $k=\frac{\pi}{a_0}$ with the wavefunction $\phi_{R\uparrow}$
localized on the right zigzag edge.  
For a short-ranged potential, 
a significant wave vector transfer  in a backscattering  occurs  for $|k - k'|\sim 1/a_0$ \cite{Lima}.  Such a short-ranged disorder potential   couples the {\it chiral}
zigzag edge state $\phi_{R\uparrow}$ to another {\it chiral} zigzag edge state $\phi_{L\uparrow}$ on the opposite zigzag edge at    $k= -\frac{\pi}{a_0}$, as shown in  Fig.\ref{Ubandst}  ($\phi_{R}$   and  $\phi_{L}$ are depicted in Fig.\ref{av1}(a)).
This process produces the bonding 
$\frac{1}{\sqrt{2}} (\phi_L +\phi_R )$ or antibonding  $\frac{1}{\sqrt{2}} (\phi_L -\phi_R )$ edge state.  These states display charge fractionalization with $1/2$ charges on the left and right zigzag edges.  The probability density of one of these  states is shown schematically in Fig.\ref{av1}(b) (a mixed chiral state).  
Moreover,   when the disorder is weak, the DOS near the gap edges  is {\it sharply peaked}.  
Because of this  sharp peak,   even a weak  disorder potential can mix the   left and right zigzag edge   states   and may generate edge states that are  fractionalized between the opposite zigzag edges.  
These results suggest that solitonic fractional charges  may exist in disordered  ZGNRs.

The midgap states with   $E\approx 0$   are of special interest as they  are affected most significantly among the gap states.
Note that the presence of the midgap states changes  the magnetic properties of ZGNRs:   the edge antiferromagnetism is weakened with the proliferation of  midgap states.  A self-consistent  treatment of disorder and electron interaction within the  HFA \cite{Mac} shows that a midgap state  of a given spin value can fractionalize into   two $e/2$  fractional boundary charges  on the opposite zigzag edges (see Fig.\ref{soliton} (e)  and Fig.\ref{av1}(b)).  
This state is divided into two equal parts and  decays
exponentially from the zigzag edges \cite{Neto}.  If an electron is added   to  a midgap  state, the resulting total
electron density  $\rho(x,y)$ has fractional charges on the zigzag edges.  The
wider the distance between the opposite zigzag edges, the better the fractional quantization, as the overlap
between the fractional charges on the left and right zigzag edges decreases.   This state represents a pseudospin kink connecting the left and right zigzag edges of different chiralities. Numerical calculations show that the charge fluctuations induced by disorder  around the mean value $e/2$ are small in the weak-disorder regime \cite{Jeong2}.  However, these fluctuations increase as the strength and range of the disorder potential increase.    To constitute true charge fractionalization, the quantum charge fluctuations should occur at high frequencies.  According to Girvin \cite{Girvin},  the characteristic time scale for the charge fluctuations is inversely proportional to the relevant excitation gap.  There is a small gap  between the occupied and unoccupied midgap states in the weak-disorder regime, and this gap  induces very small time scales for quantum fluctuations \cite{Jeong2}. It is possible for charge fractionalization to  also occur on the zigzag edges of a long disordered AGNR (Fig.\ref{arm} (b)).

Other gap states in the energy interval    $[E-\delta E,E+\delta E]$  with $E\neq 0$ also have an average fractional boundary charge value of $e/2$:  half  the  states  in this interval are more localized   on the left zigzag edge, with the other half being more localized on the right zigzag edge. Thus,  their average  boundary charge on one zigzag edge is  $e/2$.   However, they have larger charge fluctuations induced by disorder in comparison to the midgap states.

\section{Summary and conclusions}

Polyacetylene and GRNs   exhibit     bulk-edge correspondence between the Zak phase and the existence of chiral boundary states; this is guaranteed by  their chiral symmetry. In this paper, various  chiral and mixed-chiral edge  modes of polyacetylene and GNRs have been described.   Mixed chiral states, formed by bonding and antibonding linear combinations of the chiral edge modes, play an important role in charge fractionalization.   	Weak disorder stabilizes soliton states with fractional edge charges in ZGNRs, in contrast to what is usually expected.

Polyacetylene and GNRs   also have several  different topological properties.    In polyacetylene, a domain-wall soliton has unusual spin and charge relations. 
Moreover, when the  intra cell hopping parameter is smaller than the inter cell hopping parameter, the  polyacetylene is topologically non-trivial 
and fractional boundary charges  exist
with a charge of $e/2$. 
In the opposite scenario, however, the  polyacetylene  is topologically trivial with zero boundary charge.  In other words, long-bond polyacetylene is topologically non-trivial, but short-bond polyacetylene is not.  
This reflects the fact that the bulk polarization depends on the choice of unit cell \cite{Van1}.   A topological phase transition exists when the intra and inter cell tunneling coefficients are equal.  

Domain-wall solitons can also exist  in GNRs.    Unusual spin and charge relations are not present in  the GNRs as a domain wall supports an even number of solitons (manifestation of "fermion doubling"); this is in contrast to the case of polyacetylene.    Moreover, fractional boundary charges do not exist in disorder-free  rectangular AGNRs and ZGNRs because of the antiferromagnetic coupling between the well-separated zigzag edges; i.e., only integer boundary charges can exist on the zigzag edges.      The choice of  unit cell is immaterial for GNRs, in contrast to polyacetylene.
Disorder has profound effects on  the zigzag edge states of GNRs.  In the presence of disorder, a half-filled  ZGNR  becomes a Mott-Anderson insulator with numerous spin-split states.     Moreover,  the gap states are localized  whereas    the states outside the gap are  delocalized.   
 Additionally, a non-trivial interplay between disorder and electron interactions induces  formation of a fractional boundary charge.    A   disorder potential, especially a short-ranged  potential, is effective in  partly mitigating  the effect of antiferromagnetic coupling and can induce formation  of  a fractional charge of midgap states.      
 Other gap states in a small  energy interval also have $e/2$ average fractional charge, but with larger charge fluctuations.   These gap-edge states represent topological kinks. Disorder thus changes topological properties of ZGNRs.   Note that, in the case of ZGNRs, boundary charges exist on the side edges of the ribbon (one at each side).   

A fractional boundary charge may be observed by adding or removing an electron from the midgap states.  Scanning tunneling microscopy \cite{And} may provide rich information on the values and fluctuations of  the boundary charges of localized gap states.   
It may be worthwhile to develope  a field theoretical description of charge fractionalization in interacting disordered ZGNRs.
 A particle with a fractional charge is usually an anyon \cite{Wil}.  More theoretical work is needed to establish this aspect, as detection of a possible anyon state in GNRs would be most interesting.

\section*{Acknowledgments}
This research was supported by the Basic Science Research Program
through the National Research Foundation of Korea (NRF), funded by the
Ministry of Education, ICT $\&$ Future Planning (MSIP) (NRF-2018R1D1A1A09082332).


\begin{thebibliography}{9}
\bibitem{pseudo}   Consider a graphene sublattice basis consisting of two carbon atoms A and B.  Suppose that the probability amplitudes of finding an electron on the A and B atoms are $c_1$ and $c_2$, respectively.   
This scenario can be represented by a spinor $\left(\begin{array}{c}
c_1\\
c_2
\end{array}
\right)$.  Using this spinor, the components of the pseudospin  are defined as  the expectation values of the Pauli spin matrices $\sigma_x,\sigma_y$ , and $\sigma_z$.
\bibitem{kink} A kink is a one-dimensional topological object with a phase change of $\pi$ when the position coordinate varies through the kink (the crucial feature is that the manner in which the phase changes with the changing coordinate  is irrelevant).   In our case, the phase is the angle of the pseudospin vector.
\end{thebibliography}
\end{document}